\documentclass[10pt, a4paper]{article}

\usepackage[final]{lrec2026}

\usepackage{fontspec}
\usepackage{polyglossia}

\setmainlanguage{english}
\setotherlanguages{russian,korean}

\newfontfamily\russianfont[
    Path        = fonts/,
    Extension   = .otf,
    Script      = Cyrillic,
    UprightFont = TexgyrecursorRegular,
    BoldFont = TexgyrecursorBold,
    ItalicFont = TexgyrecursorItalic,
    BoldItalicFont = TexgyrecursorBolditalic
]{tex_gyre_cursor_for_russian}

\newfontfamily\koreanfont[
    Path        = fonts/,
    Extension   = .ttf,
    UprightFont = NotoSansKRThin,
]{noto_sans_cjk_kr}

\usepackage{makecell}
\usepackage{diagbox}
\usepackage{array}
\usepackage{acro}
\usepackage{arydshln}
\usepackage{tabularx}
\usepackage{hyperref}

\usepackage{anyfontsize}
\usepackage{graphicx}
\usepackage{booktabs}
\usepackage{multirow}
\usepackage{colortbl}
\usepackage{xcolor}
\usepackage{rotating}
\usepackage{tikz}

\DeclareAcronym{gpu}{
 short = GPU,
 long = graphics processing unit,
}

\DeclareAcronym{cvc}{
 short = CVC,
 long = Common Voice Corpus 17.0,
}

\DeclareAcronym{kaznerd}{
	short = KazNERD,
	long = Kazakh Named Entity Recognition Dataset
}

\DeclareAcronym{ner}{
	short = NER,
	long = named entity recognition
}

\DeclareAcronym{nlp}{
	short = NLP,
	long = natural language processing
}

\DeclareAcronym{asr}{
	short = ASR,
	long = automatic speech recognition
}

\DeclareAcronym{sota}{
	short = SOTA,
	long = state-of-the-art
}

\DeclareAcronym{tts}{
	short = TTS,
	long = text-to-speech
}

\DeclareAcronym{wer}{
	short = WER,
	long = word error rate
}

\DeclareAcronym{cer}{
	short = CER,
	long = character error rate
}

\DeclareAcronym{ksc}{
 short = KSC2,
 long = Kazakh Speech Corpus 2
}

\DeclareAcronym{kazakhtts}{
 short = KazakhTTS,
 long = Kazakh Text-to-Speech
}

\DeclareAcronym{kazakhtts2}{
 short = KazakhTTS2,
 long = Kazakh Text-to-Speech 2
}

\DeclareAcronym{kazemotts}{
 short = KazEmoTTS,
 long = Kazakh Emotional Text-to-Speech Synthesis
}

\DeclareAcronym{fleurs}{
 short = FLEURS,
 long = Few-shot Learning Evaluation of Universal Representations of Speech
}

\DeclareAcronym{nes}{
 short = NEs,
 long = named entities
}

\title{Using Songs to Improve Kazakh Automatic Speech Recognition}

\name{Rustem Yeshpanov} 

\address{Independent Researcher\\
        Astana, Kazakhstan \\
        yeshpanov.rustem@gmail.com}

\abstract{
Developing automatic speech recognition (ASR) systems for low-resource languages is hindered by the scarcity of transcribed corpora. This proof-of-concept study explores songs as an unconventional yet promising data source for Kazakh ASR. We curate a dataset of 3,013 audio-text pairs (about 4.5 hours) from 195 songs by 36 artists, segmented at the lyric-line level. Using Whisper as the base recogniser, we fine-tune models under seven training scenarios involving Songs, Common Voice Corpus (CVC), and FLEURS, and evaluate them on three benchmarks: CVC, FLEURS, and Kazakh Speech Corpus 2 (KSC2).
Results show that song-based fine-tuning improves performance over zero-shot baselines. For instance, Whisper Large-V3 Turbo trained on a mixture of Songs, CVC, and FLEURS achieves 27.6\% normalised WER on CVC and 11.8\% on FLEURS, while halving the error on KSC2 (39.3\% vs. 81.2\%) relative to the zero-shot model. Although these gains remain below those of models trained on the 1,100-hour KSC2 corpus, they demonstrate that even modest song-speech mixtures can yield meaningful adaptation improvements in low-resource ASR.
The dataset is released on Hugging Face for research purposes under a gated, non-commercial licence.
\\ \newline  }

\begin{document}

\maketitleabstract

\section{Introduction}

\Ac{asr} systems convert speech into text and typically rely on large, diverse, and carefully transcribed corpora to achieve robust performance. Such resources remain scarce for most languages worldwide. Although recent work has improved the situation for Kazakh (a Turkic language spoken by more than 15 million people worldwide), it still qualifies as low-resource by global standards, while several other Turkic languages remain even more under-resourced~\cite{veitsman-hartmann-2025-recent}.

We hypothesise that songs offer a practical, widely available, and language-agnostic data source that can complement or partially substitute conventional speech corpora for low-resource \ac{asr}. Songs are prevalent across many languages and communities; their recordings are often high quality, and accompanying lyrics can serve as approximate transcriptions. At the same time, songs pose challenges for \ac{asr}—background music, non-conversational prosody, elongated vowels, and repetitions—which makes their net utility an empirical question. This paper presents a feasibility study that asks: \emph{To what extent can song audio with aligned lyric segments help adapt \ac{asr} for Kazakh?}

To investigate this, we curated a small song-based dataset of Kazakh consisting of 3{,}013 audio-text pairs drawn from 195 songs performed by 36 artists, amounting to approximately 4.5 hours. Using Whisper~\cite{whisper} as the base recogniser, we compare zero-shot models against fine-tuned models trained on songs, on each small corpus individually (\ac{cvc}~\cite{cvc}, \ac{fleurs}~\cite{fleurs}), on each pair, and on all three combined, to systematically assess where songs help. For context, we also compare against an ``upper bound'' scenario that assumes access to over 1,100 hours of transcribed recordings from \acf{ksc}~\cite{ksc2}, a scale of data far beyond what most low-resource languages can provide.

Our experiments show that incorporating songs yields substantial improvements over zero-shot baselines on multiple benchmarks (notably on \ac{cvc} and \ac{fleurs}), and that pairing songs with small speech corpora further strengthens generalisation. However, the gains are uneven across evaluation domains, and song-augmented models do not match the performance of the large-corpus upper bound trained on \ac{ksc}. These findings suggest that songs are not a panacea, but they are a promising, broadly accessible resource that can provide measurable benefits when large speech corpora are unavailable.

The remainder of the paper is organised as follows. Section~\ref{related_work} reviews related work on low-resource \ac{asr}. Section~\ref{dataset} details the construction of the song dataset. Section~\ref{methodology} describes the experimental setup. Section~\ref{results} presents results and analysis, and Section~\ref{conclusion} concludes with limitations and directions for future work.

\raggedbottom

\section{Related Work}\label{related_work}

Research on Kazakh \ac{asr} has accelerated in recent years, supported by several initiatives that created foundational resources. \ac{ksc}~\cite{ksc2}, for example, provides 1,128 hours of transcribed audio across more than 520,000 utterances. Other efforts, such as \acs{kazakhtts}~\cite{kazakhtts}, its extension \acs{kazakhtts2}~\cite{kazakhtts2}, and \acs{kazemotts}~\cite{kazemotts}, have produced high-quality text-to-speech datasets that also support \ac{asr} research. Despite these contributions, Kazakh resources remain small compared to those available for high-resource languages, highlighting the need for alternative data sources.

Open-source multilingual datasets such as \ac{cvc}~\cite{cvc} and \ac{fleurs}~\cite{fleurs} include Kazakh and have proven valuable for training and benchmarking. Yet their scale and diversity are modest relative to corpora in languages like English or Mandarin~\cite{parcollet25_interspeech,Zhang2021WENETSPEECHA1}, limiting their effectiveness for robust \ac{asr} development.

To address such scarcity, prior work in low-resource \ac{asr} has focused on techniques such as data augmentation and transfer learning. Data augmentation has been shown to improve performance by synthesising additional speech data~\cite{baas22_interspeech,casanova23_interspeech}, while transfer learning approaches leverage transliteration~\cite{khare21_interspeech} or cross-lingual speech-to-text translation~\cite{wang20ia_interspeech} to adapt models from high-resource languages. These methods, however, often still rely on a non-trivial amount of initial speech data, making them less applicable in scenarios where only minimal resources are available.

This motivates exploration of alternative resources that are both accessible and linguistically broad. In this paper, we propose the use of songs as such a resource. Songs exist in nearly every language, are often distributed in high-quality recordings, and are accompanied by lyrics that provide approximate transcriptions. While challenges include background music and non-standard prosody, their universality makes songs a promising but underexplored candidate for low-resource \ac{asr}. To our knowledge, no prior work has systematically investigated the use of songs for Kazakh or other low-resource languages, making this study a first step in addressing this gap.

\section{Dataset}\label{dataset}

\subsection{Song Selection and Collection}

The dataset was collected over four months by the author. To achieve diversity, 195 Kazakh songs performed by 36 artists (14 female and 22 male) were selected. Songs were downloaded from YouTube using the \texttt{youtube\_dl} library\footnote{\url{https://pypi.org/project/youtube_dl/}}. Songs were selected according to the following criteria:

\begin{itemize}
    \item \textbf{Vocal focus:} Only tracks featuring prominent solo vocals were included; songs dominated by choral arrangements, duets/overlapping lead vocals, or instrumental-only sections were excluded.
    \item \textbf{Genre diversity:} Songs representing a range of mainstream Kazakh music were included to capture stylistic variety (pop, pop-estrada, folk-pop, folk-rock, R\&B, and hip-hop).
    \item \textbf{Artist representation:} Popular contemporary artists with publicly available songs were included.
    \item \textbf{Source search strategy:} Songs were identified using artist discographies, popular streaming platforms, and online music catalogues. Keywords included artist names, album titles, and widely recognised track titles.
\end{itemize}

In the interest of space, we do not list all artists and songs; however, each audio-text pair retains metadata with the corresponding artist name and song title to support reproducibility and future reference. In addition, each segment retains start/end timestamps to make the line-level segmentation reproducible.

\subsection{Data Preparation and Validation}

Preprocessing involved several steps to prepare high-quality audio-text pairs suitable for \ac{asr}. First, vocal tracks were separated from instrumental accompaniment using \texttt{Spleeter}\footnote{\url{https://pypi.org/project/spleeter/}}. While this process does not perfectly isolate vocals and may leave residual background music, the resulting tracks were deemed sufficiently clear to preserve intelligible speech content. As a quality-control step, we manually audited the separated vocals by listening and discarded segments where the lyrics were not intelligible due to residual accompaniment or separation artefacts.

Lyrics were collected from online repositories and official artist websites. As the accuracy of these sources varied, the lyrics were manually reviewed and corrected to match the actual sung content, including repetitions and colloquial pronunciations. Corrected texts retained original casing and punctuation, reflecting the structure of the lyrics rather than applying \ac{asr}-style normalisation.

Alignment was performed manually using \texttt{Audacity}\footnote{\url{https://www.audacityteam.org/}}, an open-source audio editing tool. Each song was segmented at the line level by listening and synchronising with the corrected lyrics, after which the “Export Labels” feature in \texttt{Audacity} was used to generate audio-text pairs.

The resulting dataset comprises 3,013 pairs with a total duration of approximately 4.5 hours. Summary statistics are presented in Table~\ref{tab:summary_stats}. Note that the totals for unique utterances and words are lower than the sum of female and male counts, as some material occurs in songs by both groups; such overlaps were removed in the overall totals. Mean utterance length was nearly identical across male and female artists (5.4 s), an outcome that appears coincidental given the diversity of genres and performers. The dataset, while carefully constructed, also presents several limitations, discussed below.

\begin{table}[ht]
\centering
\begin{tabular}{rrrr}
\toprule
 & \textbf{Female} & \textbf{Male} & \textbf{Total} \\
 \midrule
 \textbf{Artists} & 14 & 22 & 36 \\
\textbf{Utterances} & 1,387 & 1,626 & 3,013 \\
\textbf{Unique utterances} & 1,369 & 1,598 & 2,938 \\
\textbf{Words} & 6,504 & 7,441 & 13,945 \\
\textbf{Unique words} & 3,075 & 3,388 & 5,359 \\
\textbf{Duration (h)} & 2.1 & 2.4 & 4.5 \\
\textbf{Min (s)} & 1.0 & 1.1 & 1.0 \\
\textbf{Max (s)} & 17.0 & 15.3 & 17.0 \\
\textbf{Mean (s)} & 5.4 & 5.4 & 5.4 \\
\bottomrule
\end{tabular}
\caption{Summary statistics of the Kazakh songs dataset \label{tab:summary_stats}}
\end{table}

\subsection*{Limitations}

As with any dataset of this scale and nature, it has several limitations. First, its total duration of 4.5 hours is relatively small compared to conventional \ac{asr} training corpora, which may restrict model generalisation. Second, although multiple genres were included, the selection is not exhaustive and may underrepresent less common styles or regional variations. Third, despite vocal separation with \texttt{Spleeter}, residual background music remains in some segments, potentially introducing noise. Finally, alignment and lyric correction were performed manually by the author, which, although carefully conducted, may introduce subjective inconsistencies. These limitations highlight that the dataset should be viewed as a proof-of-concept resource rather than a comprehensive corpus.

\section{Methodology}\label{methodology}

\subsection{Training Setup}

Given the computational expense of fine-tuning large \ac{asr} models, all experiments were conducted on \texttt{Vast.AI}\footnote{\url{https://vast.ai/}} using an NVIDIA RTX 3090 GPU (24~GB VRAM). Across all fine-tuning runs, the total compute cost was approximately \$25, underscoring the affordability of this approach relative to large-scale training.

The fine-tuning configuration was kept consistent across experiments. We used an initial learning rate of $5 \times 10^{-6}$ with 50 warm-up steps, a batch size of 60, and an early stopping criterion with a patience of two epochs. Training was conducted under seven scenarios using the training splits of the respective datasets: (i) Songs only, (ii) \ac{cvc} only, (iii) \ac{fleurs} only, (iv) Songs + \ac{cvc}, (v) Songs + \ac{fleurs}, (vi) \ac{cvc} + \ac{fleurs}, and (vii) Songs + \ac{cvc} + \ac{fleurs}. Validation was performed on the \ac{cvc} and \ac{fleurs} validation sets. This choice reflects the fact that the Songs dataset was intended solely as a training resource, and that the \ac{ksc} corpus differs substantially in format (all lowercase, no punctuation). In contrast, \ac{cvc} and \ac{fleurs} include casing and punctuation, making them more compatible with the song-based training data and more informative for monitoring model generalisation. These format differences underscore the broader challenge of cross-dataset evaluation in Kazakh \ac{asr}, where corpora often vary in orthographic conventions and normalisation standards. Final evaluation was conducted on three independent benchmarks: the \ac{ksc} test set~\cite{ksc2}, the \ac{cvc} test set~\cite{cvc}, and the \ac{fleurs} test set~\cite{fleurs}.

The benchmark datasets are briefly summarised below.  
\ac{ksc} contains 1,128 hours of transcribed audio, crowdsourced and collected for both \ac{asr} and \ac{tts} development. It spans a wide range of sources, including news broadcasts, radio programs, parliamentary speeches, podcasts, and Kazakh-Russian code-switching utterances.  
\ac{cvc} is a large-scale, volunteer-based corpus for multilingual \ac{asr}, which includes a Kazakh subset. Inspection of the Kazakh test set revealed that it predominantly consists of sayings and proverbs.  
\ac{fleurs} is a multilingual read-speech benchmark developed by Google Research. It extends the FLoRes machine translation dataset~\cite{flores} by adding speech recordings of its sentences, originally derived from English Wikipedia. The Kazakh subset includes some words of foreign origin spelt in Latin script, in contrast to the Cyrillic-only orthography of the other test sets used here.

Before training and evaluation, additional preprocessing was applied for consistency. Two English sentences were identified in the \ac{fleurs} training set and removed. Sentences in the \ac{ksc} test set that contained only Russian words were identified and removed. In addition, homoglyphs appearing in both the \ac{ksc} and \ac{cvc} test sets were replaced with their respective Kazakh letters to maintain script consistency.

Another important aspect of this study is that the datasets differ in their transcription conventions. The Songs dataset and \ac{cvc} both use Cyrillic script with casing and punctuation, but contain no digits. In contrast, \ac{fleurs} includes a mix of Cyrillic and Latin script, digits, casing, and punctuation. The \ac{ksc} corpus is distinct from all of these, being entirely lowercased, Cyrillic-only, and stripped of punctuation. Table~\ref{tab:formats} summarises transcription format differences across datasets, while Table~\ref{tab:splits} reports statistics for the specific splits used in this study—namely, the training, validation, and test sets of \ac{cvc} and \ac{fleurs}, and the test set only for \ac{ksc}.

\begin{table}[h]
\centering

\fontsize{8.98}{12}\selectfont
\renewcommand{\arraystretch}{1.2}
\setlength{\tabcolsep}{0.08cm}
\begin{tabular}{lcccc}
\toprule
\textbf{Dataset} & \textbf{Script} & \textbf{Casing} & \textbf{Punctuation} & \textbf{Digits} \\
\midrule
Songs   & Cyrillic & Yes & Yes & No \\
\ac{cvc}     & Cyrillic & Yes & Yes & No \\
\ac{fleurs}  & Cyrillic + Latin & Yes & Yes & Yes \\
\ac{ksc}    & Cyrillic & No  & No  & No \\
\bottomrule
\end{tabular}
\caption{\label{tab:formats} Transcription conventions across datasets}
\end{table}

\begin{table}[ht]
\centering
\fontsize{7}{11}\selectfont
\setlength\tabcolsep{0.075cm}
\begin{tabularx}{\columnwidth}{rrrrrrrr}
\toprule
 & \multicolumn{3}{c}{\textbf{\ac{cvc}}} & \multicolumn{3}{c}{\textbf{\ac{fleurs}}} & \multicolumn{1}{c}{\textbf{\ac{ksc}}} \\
 \cmidrule{2-4}  \cmidrule{5-7} \cmidrule{8-8}
 & \multicolumn{1}{c}{\textbf{Train}} & \multicolumn{1}{c}{\textbf{Valid}} & \multicolumn{1}{c}{\textbf{Test}} & \multicolumn{1}{c}{\textbf{Train}} & \multicolumn{1}{c}{\textbf{Valid}} & \multicolumn{1}{c}{\textbf{Test}} & \multicolumn{1}{c}{\textbf{Test}} \\
\midrule
Speakers & 4 & 21 & 106 & - & - & - & - \\
Utterances & 548 & 498 & 514 & 3,198 & 369 & 856 & 9,192 \\
Unique utter. & 548 & 498 & 513 & 1,493 & 147 & 349 & 9,072 \\
Words & 3,324 & 2,989 & 3,100 & 53,520 & 5,877 & 15,014 & 102,035 \\
Unique words & 1,990 & 1,784 & 1,873 & 10,358 & 1,675 & 3,502 & 23,678 \\
Duration (h) & 0.7 & 0.6 & 0.7 & 11.8 & 1.5 & 3.8 & 15.6 \\
min (s) & 2.2 & 1.9 & 2.1 & 3.0 & 4.7 & 5.8 & 1.0 \\
max (s) & 10.5 & 10.5 & 9.7 & 36.0 & 46.0 & 43.3 & 26.3 \\
mean (s) & 4.9 & 4.6 & 5.1 & 13.3 & 14.9 & 16.1 & 6.1 \\
\bottomrule
\end{tabularx}
\caption{Dataset splits and test set statistics \label{tab:splits}}
\end{table}

\subsection{Model Selection}

We first evaluated whether fine-tuning smaller models on the collected dataset yields measurable performance improvements. Specifically, we experimented with \texttt{Whisper-tiny}\footnote{\url{https://huggingface.co/openai/whisper-tiny}} (39M parameters) and \texttt{Whisper-small}\footnote{\url{https://huggingface.co/openai/whisper-small}} (244M parameters), two compact variants from OpenAI’s Whisper family~\cite{whisper}. Due to space constraints, Table~\ref{tab:first_results} reports \ac{wer} and \ac{cer} on three test sets (\ac{cvc}, \ac{fleurs}, \ac{ksc}) using normalised transcripts, which provide consistency across datasets with differing conventions. Results show that even these small-scale models exhibit clear relative improvements after fine-tuning (FT) on the Songs dataset compared to their pre-trained (PT) baselines, although absolute error rates remain high.

\begin{table}[h!]
\centering
\renewcommand{\arraystretch}{1.2}
\fontsize{8}{12}\selectfont
\setlength\tabcolsep{0.11cm}
\begin{tabularx}{\columnwidth}{ccrrcrrcrr}
\toprule
\multicolumn{2}{c}{\multirow{2}{*}{\textbf{Whisper}}} & \multicolumn{2}{c}{\textbf{\ac{cvc}}} &  & \multicolumn{2}{c}{\textbf{\ac{fleurs}}} &  & \multicolumn{2}{c}{\textbf{\ac{ksc}}} \\
\cmidrule{3-4} \cmidrule{6-7} \cmidrule{9-10}
 &  & \textbf{\ac{wer}} & \textbf{\ac{cer}} &  & \textbf{\ac{wer}} & \textbf{\ac{cer}} &  & \textbf{\ac{wer}} & \textbf{\ac{cer}} \\
\midrule
tiny & PT & 190.7 & 160.7 &  & 154.4 & 110.0 &  & 219.2 & 156.4 \\
 & FT & 87.0 & 28.4 &  & 99.8 & 31.1 &  & 111.5 & 46.5 \\
small & PT & 179.6 & 108.9 &  & 89.8 & 48.0 &  & 125.0 & 81.1 \\
 & FT & 60.7 & 16.4 &  & 63.8 & 17.4 &  & 78.5 & 27.8 \\
\bottomrule
\end{tabularx}
\caption{Normalised \ac{wer} and \ac{cer} performance of Whisper-tiny and Whisper-small \label{tab:first_results}}
\end{table}

Given the high computational cost of training Whisper Large-V3 (1550M parameters) directly, we instead adopted Whisper Large-V3 Turbo\footnote{\url{https://huggingface.co/openai/whisper-large-v3-turbo}}, a pruned variant that reduces the number of decoding layers (32 to 4) and achieves substantially faster inference with only minor quality degradation. For additional context, we also included a community fine-tuned version of Whisper Large-V3 Turbo trained on \ac{ksc}\footnote{\url{https://huggingface.co/abilmansplus/whisper-turbo-ksc2}}, representing an upper-bound scenario unlikely to be available for most low-resource languages.

\subsection{Evaluation Metrics}

To assess \ac{asr} performance, this study employed \ac{wer} and \ac{cer}, both of which are standard evaluation metrics in speech recognition~\cite{wer,cer}. \ac{wer} measures transcription accuracy at the word level, while \ac{cer} provides finer-grained evaluation at the character level, making it particularly useful for languages with rich morphology such as Kazakh.

Two versions of \ac{wer} and \ac{cer} were computed. The first, orthographic \ac{wer} and \ac{cer} (\ac{wer}$_{or}$ and \ac{cer}$_{or}$), was calculated using the original transcriptions, preserving casing, punctuation, and extra whitespace. The second, normalised \ac{wer} and \ac{cer} (\ac{wer}$_{no}$ and \ac{cer}$_{no}$), was computed after normalising text by lowercasing, removing punctuation, and collapsing extra whitespace. Given the formatting of \ac{ksc} (see Table~\ref{tab:formats}), we report only normalised \ac{wer} and \ac{cer} for the respective test set.

\begin{table*}[t]
\centering
\fontsize{9}{12}\selectfont
\renewcommand{\arraystretch}{1.2}
\setlength\tabcolsep{0.05cm}
\begin{tabular}{lcccccccccc}
\toprule
 & \multicolumn{4}{c}{\textbf{\ac{cvc}}} & \multicolumn{4}{c}{\textbf{\ac{fleurs}}} & \multicolumn{2}{c}{\textbf{\ac{ksc}}} \\
\cmidrule(lr){2-5} \cmidrule(lr){6-9} \cmidrule(lr){10-11}
\textbf{Model / Training} & \textbf{\ac{wer}$_{or}$} & \textbf{\ac{cer}$_{or}$} & \textbf{\ac{wer}$_{no}$} & \textbf{\ac{cer}$_{no}$} &
\textbf{\ac{wer}$_{or}$} & \textbf{\ac{cer}$_{or}$} & \textbf{\ac{wer}$_{no}$} & \textbf{\ac{cer}$_{no}$}  & \textbf{\ac{wer}$_{no}$} & \textbf{\ac{cer}$_{no}$} \\
\midrule
\multicolumn{11}{l}{\textit{Baselines}} \\
Whisper Large-V3 (zero-shot) & 61.2 & 22.0 & 56.5 & 20.7 & 41.1 & 9.5 & 33.1 & 7.8 & 58.9 & 23.6 \\
Whisper Large-V3 Turbo (zero-shot) & 70.6 & 28.6 & 47.7 & 23.8 & 38.0 & 9.0 & 21.0 & 6.0 & 81.2 & 42.6 \\
Community fine-tuned (\ac{ksc}) & 56.2 & 10.9 & 12.5 & 3.1 & 36.0 & 8.8 & 11.3 & 5.1 & 9.3 & 3.2 \\
\midrule
\multicolumn{11}{l}{\textit{Fine-tuning Whisper Large-V3 Turbo}} \\
+ Songs only               & 49.8 & 12.3 & 37.3 & 9.3 & 33.9 & 7.6 & 23.7 & 5.7 & 45.2 & 15.2 \\
+ \ac{cvc} only                  & 42.3 & 10.0 & 39.1 & 9.2 & 48.1 & 11.3 & 37.9 & 9.2 & 58.1 & 19.6 \\
+ \ac{fleurs} only                & 51.6 & 14.8 & 43.6 & 12.8 & 21.0 & 4.4 & 13.6 & 3.1 & 46.6 & 18.7 \\
+ Songs + \ac{cvc}            & 33.9 & 8.7  & 29.6 & 7.8 & 33.8 & 7.5 & 23.7 & 5.6 & 43.7 & 14.2 \\
+ Songs + \ac{fleurs}              & 40.9 & 11.0 & 34.1 & 9.5 & 19.7 & 3.9 & 11.8 & 2.6 & 40.4 & 16.0 \\
+ \ac{cvc} + \ac{fleurs}               & 33.0 & 7.7  & 28.1 & 6.6 & 19.7 & 4.0 & 11.8 & 2.6 & 39.3 & 13.9 \\
+ Songs + \ac{cvc} + \ac{fleurs}      & 32.0 & 7.4  & 27.6 & 6.5 & 19.5 & 3.9 & 11.8 & 2.6 & 39.3 & 14.4 \\
\midrule
\multicolumn{11}{l}{\textit{Fine-tuning Community Model (\ac{ksc})}} \\
+ Songs only                & 41.1 & 8.5  & 13.9 & 3.5 & 32.6 & 7.9 & 11.7 & 4.7 & 10.3 & 3.5 \\
+ \ac{cvc} only                  & 56.3 & 10.9 & 12.6 & 3.1 & 35.7 & 8.3 & 10.8 & 4.5 & 9.3  & 3.2 \\
+ \ac{fleurs} only               & 26.4 & 6.1  & 14.4 & 3.7 & 16.0 & 3.6 & 7.3  & 2.1 & 13.9 & 5.8 \\
+ Songs + \ac{cvc}            & 19.5 & 4.5  & 12.3 & 3.1 & 26.5 & 7.2 & 11.5 & 4.8 & 10.3 & 3.5 \\
+ Songs + \ac{fleurs}              & 26.8 & 7.9  & 17.1 & 6.0 & 16.3 & 4.1 & 8.4  & 2.7 & 15.2 & 6.2 \\
+  \ac{cvc} + \ac{fleurs}               & 19.1 & 4.7  & 13.5 & 3.5 & 15.4 & 3.4 & 7.1  & 2.0 & 13.8 & 5.7 \\
+ Songs + \ac{cvc} + \ac{fleurs}      & 21.0 & 5.2  & 15.6 & 4.1 & 15.8 & 3.7 & 7.9  & 2.3 & 16.3 & 6.6 \\
\bottomrule
\end{tabular}
\caption{Orthographic and normalised \ac{wer}/\ac{cer} (\%) on three Kazakh test sets \label{tab:main_results}}
\end{table*}

\section{Results}\label{results}

\begin{table*}[t]
\centering
\fontsize{8}{10}\selectfont
\renewcommand{\arraystretch}{1.2}
\setlength\tabcolsep{0.12cm}
\begin{tabular}{llrrrrrr}
\toprule
 & \textbf{Model / Training} & \textbf{Crowdsourced} & \textbf{Parliament} & \textbf{Podcasts} & \textbf{Radio} & \textbf{Talkshow} & \textbf{TV News} \\
\midrule
\multicolumn{8}{l}{\textit{Baselines}} \\
 & Whisper Large-V3 & 47.3 & 79.9 & 68.0 & 72.1 & 63.6 & 55.5 \\
 & Whisper Large-V3 Turbo & 30.2 & 68.5 & 166.5 & 88.9 & 164.3 & 60.5 \\
  & Community fine-tuned (\ac{ksc}) & 5.0 & 6.0 & 18.8 & 15.7 & 13.7 & 4.9 \\
\midrule
\multicolumn{8}{l}{\textit{Fine-tuning Whisper Large-V3 Turbo}} \\
 & + Songs only & 32.6 & 39.7 & 61.8 & 62.9 & 60.4 & 41.3 \\
 & + \ac{cvc} only & 44.7 & 55.3 & 73.5 & 75.7 & 68.6 & 56.6 \\
 & + \ac{fleurs} only & 35.6 & 37.3 & 64.5 & 67.3 & 59.3 & 40.1 \\
 & + Songs + \ac{cvc} & 31.5 & 42.7 & 59.4 & 60.8 & 52.1 & 40.3 \\
 & + Songs + \ac{fleurs} & 27.7 & 33.3 & 59.0 & 58.3 & 56.4 & 35.2 \\
 & + \ac{cvc} + \ac{fleurs} & 27.2 & 37.5 & 55.5 & 59.2 & 51.1 & 33.9 \\
 & + Songs + \ac{cvc} + \ac{fleurs} & 28.0 & 34.6 & 55.7 & 57.1 & 49.9 & 34.6 \\
\midrule
\multicolumn{8}{l}{\textit{Fine-tuning Community Model (\ac{ksc})}} \\
 & + Songs only & 6.1 & 6.8 & 20.6 & 15.8 & 15.7 & 5.4 \\
 & + \ac{cvc} only & 5.1 & 5.9 & 18.8 & 15.5 & 13.8 & 4.9\\
 & + \ac{fleurs} only & 12.8 & 9.1 & 20.8 & 16.5 & 15.4 & 9.3 \\
 & + Songs + \ac{cvc} & 6.0 & 6.6 & 20.1 & 16.5 & 15.5 & 5.5\\
 & + Songs + \ac{fleurs} & 13.4 & 10.6 & 23.2 & 19.6 & 17.2 & 10.2 \\
 & + \ac{cvc} + \ac{fleurs} & 12.5 & 8.9 & 20.7 & 17.1 & 15.9 & 9.4 \\
 & + Songs + \ac{cvc} + \ac{fleurs} & 13.9 & 11.8 & 25.3 & 21.9 & 18.5 & 10.9 \\
\bottomrule
\end{tabular}
\caption{Normalised \ac{wer} (\%) across six speech domains of the \ac{ksc} test set \label{tab:ksc_domains}}
\end{table*}

\subsection{Quantitative Evaluation}

\subsubsection{Baselines}

Table~\ref{tab:main_results} reports performance on three evaluation sets (\ac{cvc}, \ac{fleurs}, \ac{ksc}). Zero-shot models struggle on Kazakh overall. On \ac{cvc} (normalised), Whisper Large-V3 achieves 56.5, while the pruned Turbo variant is better at 47.7; however, Turbo is markedly worse on \ac{ksc} (81.2 vs.\ 58.9). Orthographic scores on \ac{cvc} show the opposite trend: Turbo degrades from 61.2 to 70.6, reflecting its difficulty with casing and punctuation. In contrast, the community fine-tuned model trained on the 1{,}128-hour \ac{ksc} corpus establishes a strong upper bound with 12.5 (\ac{cvc}), 11.3 (\ac{fleurs}), and 9.3 (\ac{ksc}) normalised \ac{wer}.

\subsubsection{Fine-tuning Whisper Large-V3 Turbo}

We fine-tuned the Turbo model on Songs, \ac{cvc}, and \ac{fleurs}—individually and in mixtures.

\paragraph{Single-source fine-tuning.}
\emph{Songs only} improves \ac{cvc} and \ac{ksc} over the Turbo baseline (\ac{cvc}: 37.3 vs.\ 47.7; \ac{ksc}: 45.2 vs.\ 81.2 normalised \ac{wer}), but hurts \ac{fleurs} (23.7 vs.\ 21.0). 
\emph{\ac{cvc} only} helps in-domain (39.1 on \ac{cvc}) but remains weak out-of-domain (\ac{fleurs} 37.9; \ac{ksc} 58.1). 
\emph{\ac{fleurs} only} helps \ac{fleurs} strongly (13.6) and also improves \ac{ksc} vs.\ baseline (46.6), with modest \ac{cvc} gains (43.6).

\paragraph{Mixtures.}
Mixtures are consistently stronger than single sources. 
\emph{\ac{cvc} + \ac{fleurs}} yields 28.1 (\ac{cvc}) and 11.8 (\ac{fleurs}) normalised \ac{wer}, and improves \ac{ksc} to 39.3. 
The triple mixture (\emph{Songs + \ac{cvc} + \ac{fleurs}}) is the most balanced overall: 27.6 on \ac{cvc} (best), 11.8 on \ac{fleurs} (ties best), and 39.3 on \ac{ksc} (ties best). 
\emph{Songs + \ac{fleurs}} also reaches 11.8 on \ac{fleurs} but is weaker on \ac{cvc} (34.1) and \ac{ksc} (40.4); 
\emph{Songs + \ac{cvc}} prioritises \ac{cvc} (29.6) but leaves \ac{fleurs} relatively high (23.7).

\paragraph{Orthographic view.}
Orthographic improvements mirror the normalised trends. 
The triple mixture reduces orthographic \ac{wer} on \ac{cvc} from 70.6 (Turbo baseline) to 32.0 (a \(\sim\)55\% relative reduction), and on \ac{fleurs} from 38.0 to 19.5 (\(\sim\)49\% relative reduction). 

\subsubsection{Fine-tuning the Community (\ac{ksc}) Model}

Starting from the \ac{ksc}-trained upper bound leaves limited headroom and introduces domain-drift risks.

\paragraph{Single-source fine-tuning.}
\emph{Songs only} degrades performance relative to the \ac{ksc} baseline (\ac{cvc} 13.9 vs.\ 12.5; \ac{fleurs} 11.7 vs.\ 11.3; \ac{ksc} 10.3 vs.\ 9.3). 
\emph{\ac{cvc} only} largely preserves \ac{ksc} (9.3) and slightly improves \ac{fleurs} (10.8), with \ac{cvc} roughly unchanged (12.6). 
\emph{\ac{fleurs} only} achieves a large gain on \ac{fleurs} (7.3) but substantially harms \ac{ksc} (13.9) and worsens \ac{cvc} (14.4).

\begin{table*}[!ht]
\centering
\fontsize{6}{8}\selectfont
\begin{center}
\begin{tabular}{cclr}
\toprule
\multicolumn{1}{c}{\textbf{System output}} & \multicolumn{1}{c}{\textbf{Test set}} & \multicolumn{1}{c}{\textbf{Text}} & \multicolumn{1}{c}{\textbf{WER}} \\
\midrule
Reference & \multirow{6}{*}{CVC} &  \multicolumn{1}{l}{\makecell[l]{\textit{\textrussian{Жақсыда жаттық жоқ, жаманда достық жоқ.}} \\ Zhaqsyda zhattyq zhoq, zhamanda dostyq zhoq. \\ The good have no strangers; the bad have no friends.}} & 0.0 \\ \cdashline{3-4}
Whisper Large-V3 & & \textrussian{Şahsı da jaktozok, samanda toztozok.} & 100.0 \\
Whisper Large-V3 Turbo & & \textkorean{자수다자 도적 삼엔더 도적} & 100.0 \\
Community fine-tuned & & \textrussian{жақсы да жаттық жоқ жаман да достық жоқ} & 100.0 \\
WLT\_SCF & & \textrussian{Жақсы да жаттыұ жоқ, жаман да достыұ жоқ.} & 100.0 \\
CFT\_SCF & & \textrussian{Жақсыда жаттық жоқ, жаманда достық жоқ.} & 0.0 \\
\midrule
Reference & \multirow{6}{*}{FLEURS} & \multicolumn{1}{l}{\makecell[l]{\textit{\textrussian{Матаның тым ыстық болуына жол бермеңіз (бұл қысқаруға немесе күюіне себеп болуы мүмкін).}} \\ Matanyng tym ystyq boluyna zhol bermengiz (būl qysqaruğa nemese küyüine sebep boluy mümkin). \\ Do not expose the fabric to excessive heat (this may cause shrinkage or scorching).}} & 0.0 \\ \cdashline{3-4}
Whisper Large-V3 & & \textrussian{Матаның тұм ұстық болуына жол берменіз. Бұл қысқаруа немесе күйіне себеп болуы мүмкін.} & 53.85 \\
Whisper Large-V3 Turbo & & \textrussian{матаның тым ыстық болуына жол бермеңіз бұл қысқаруға немесе күйеуіне себеп болуы мүмкін} & 30.77 \\
Community fine-tuned & & \textrussian{матаның тым ыстық болуына жол бермеңіз бұл қысқаруға немесе күюіне себеп болуы мүмкін} & 23.08 \\
WLT\_SCF & & \textrussian{Матаның тым ыстық болуына жол бермеңіз. Бұл қысқаруа немесе күюіне себеп болуы мүмкін.} & 30.77 \\
CFT\_SCF & & \textrussian{Матаның тым ыстық болуына жол бермеңіз. Бұл қысқаруға немесе күюіне себеп болуы мүмкін.} & 23.08 \\
\midrule
Reference & \multirow{6}{*}{KSC2} & \multicolumn{1}{l}{\makecell[l]{\textit{\textrussian{осыны мəселені есте ұстауымыз керек}} \\ osyny mäseleni este ūstauymyz kerek \\ We should keep this issue in mind.}} & 0.0 \\ \cdashline{3-4}
Whisper Large-V3 & & \textrussian{основным осиленным из тех стал маскерек} & 120.0 \\
Whisper Large-V3 Turbo & & \textrussian{osnamaðsýlina í stíðstáum skýrek} & 100.0 \\
Community fine-tuned & & \textrussian{осындай мәселені есте ұстауымыз керек} & 20.0 \\
WLT\_SCF & & \textrussian{осыны мәселені есте ұстауымыз керек} & 0.0 \\
CFT\_SCF & & \textrussian{осындай мәселені есте ұстауымыз керек} & 20.0 \\
\bottomrule
\end{tabular}
\end{center}
\caption{Sample \ac{asr} outputs across \ac{cvc}, \ac{fleurs}, and \ac{ksc} for zero-shot, community fine-tuned, and song-adapted systems (WLT\_SCF and CFT\_SCF). WER$_{or}$ is reported for \ac{cvc} and \ac{fleurs}, and WER$_{no}$ for \ac{ksc}.\label{tab:asr_samples}}
\end{table*}

\paragraph{Mixtures.}
\emph{\ac{cvc} + \ac{fleurs}} produces the best \ac{fleurs} normalised \ac{wer} 7.1, but with notable forgetting on \ac{ksc} (13.8). 
\emph{Songs + \ac{cvc}} keeps \ac{cvc} near the baseline (12.3) while slightly worsening \ac{fleurs} (11.5) and \ac{ksc} (10.3). 
The triple mixture (15.6 on \ac{cvc}; 7.9 on \ac{fleurs}; 16.3 on \ac{ksc}) is more balanced than \emph{\ac{fleurs} only} but still shows clear degradation on \ac{ksc} relative to the baseline.

To further assess cross-domain generalisation, Table~\ref{tab:ksc_domains} reports normalised \ac{wer} across six speech domains of the \ac{ksc} test set: crowdsourced, parliamentary, podcasts, radio, talkshows, and television news.
The largest relative gains are observed for the \texttt{Whisper Large-V3 Turbo} model fine-tuned within this study—rather than for the community model already trained on the full 1,100-hour \ac{ksc} corpus.
Song-based fine-tuning yields striking improvements in spontaneous and conversational domains such as \textit{podcasts} and \textit{talkshows}, where error rates drop by roughly two-thirds compared to the zero-shot Turbo baseline, and by about half in \textit{parliamentary} speech.
More moderate but still consistent reductions are seen for \textit{radio} and \textit{TV news}, indicating that musical data can aid adaptation even across differing acoustic and stylistic conditions. 

\subsection{Qualitative Error Analysis}

Beyond quantitative metrics, we analyse representative outputs in Table~\ref{tab:asr_samples} to understand how song-based adaptation affects recognition behaviour. Across all test sets, models further fine-tuned with Songs---Whisper Large-V3 Turbo fine-tuned on Songs + \ac{cvc} + \ac{fleurs} (WLT\_SCF) and the community \ac{ksc}-trained model further fine-tuned on Songs + \ac{cvc} + \ac{fleurs} (CFT\_SCF)---exhibit more stable and linguistically coherent transcriptions than the zero-shot baselines.

A key difference is reduced cross-lingual drift. On the \ac{ksc} sample, the zero-shot Whisper Large-V3 output shifts into another language and the turbo variant produces nonsensical tokens, whereas the song-adapted models remain consistently in Kazakh and recover the intended meaning with only minor variation. This suggests that exposure to song data strengthens lexical grounding and decoding stability under acoustically challenging conditions.

Improvements are also evident in lexical and morphological accuracy. On the \ac{cvc} example, zero-shot models produce unintelligible outputs, while song-adapted models recover the syntactic structure and core vocabulary, with only minor phonetic substitutions (e.g., \textrussian{қ} → \textrussian{ұ}) that do not affect lexical interpretability. On the \ac{fleurs} sample, which represents instructional prose, song-adapted models more reliably preserve key lexical items and suffixes, whereas zero-shot variants exhibit vowel distortions and incorrect substitutions.

Finally, qualitative differences appear in punctuation and sentence segmentation. Song-adapted models more consistently restore clause boundaries and punctuation on the \ac{fleurs} example, suggesting improved modelling of prosodic and syntactic cues. This behaviour is consistent with the nature of song data, where lyrical phrasing and rhythmic pauses provide additional boundary information.

Overall, the qualitative evidence indicates that song-based fine-tuning improves not only acoustic robustness but also language stability, lexical recovery, morphological accuracy, and resistance to cross-lingual hallucinations, complementing the quantitative \ac{wer} improvements.

\subsection{Discussion}

Three conclusions emerge. 
First, songs alone are not sufficient and may hurt some domains (e.g., \ac{fleurs} on Turbo; most domains on the \ac{ksc} model). 
Second, songs are valuable when combined with modest corpora: the triple mixture achieves the best \ac{cvc} (27.6) and ties the best \ac{fleurs} (11.8) on Turbo, while also halving \ac{ksc} error relative to the Turbo baseline (39.3 vs.\ 81.2). 
Third, once a model is trained on approximately 1,100 hours (\ac{ksc} upper bound), additional fine-tuning on songs/small corpora confers marginal gains at best and often induces forgetting.

A further factor likely contributing to the modest gains from song-only fine-tuning is dataset size.
With only 4.5 hours of audio and roughly 3,000 lyric lines, the song corpus represents a very small fraction of the data typically required to meaningfully adapt a billion-parameter model such as Whisper.
At this scale, the training signal may have been too limited to shift model parameters substantially, even though it proved complementary when combined with other small corpora.
Future work should investigate whether larger and more diverse song collections—potentially including synthetic data—could yield stronger and more consistent adaptation effects.

Orthographic metrics reinforce these findings: while casing and punctuation remain challenging, mixtures (especially those including \ac{fleurs}) deliver substantial orthographic error reductions, indicating that the benefits of songs extend beyond pure lexical recognition to improved written-form fidelity.

\vspace{-0.2cm}

\paragraph{Ethical and legal considerations.}
Beyond these technical outcomes, it is important to acknowledge the elephant in the room: the copyrighted nature of the Songs dataset. The recordings used here are copyrighted works, and no explicit permission was obtained from the artists for use in \ac{asr} development. This raises a broader question: \textit{Is the absence of prior research on songs as an \ac{asr} resource due primarily to lack of exploration, or to the legal and ethical complexities surrounding their use?}
This study is exploratory and not intended as a deployment-ready approach; rather, it seeks to assess whether songs have technical merit as a training signal. If the answer is yes, the next step would involve dialogue on how such data might be ethically and legally integrated into \ac{asr} development pipelines for low-resource languages---for example, through short excerpts, public-domain materials, collaborations with artists, or structured fair-use frameworks.

\vspace{-0.2cm}

\paragraph{Synthetic alternatives.}
A promising direction for addressing copyright concerns lies in synthetic music generation. Modern tools such as \texttt{Suno.com}\footnote{\url{https://suno.com/}} can generate songs with customisable parameters: lyrics in low-resource languages, stylistic control (e.g., folk, pop, rap), and varied vocal timbres (male/female, solo/chorus). If song-based training proves beneficial, synthetic songs could provide a scalable and legally permissible alternative. They would allow researchers to systematically generate datasets reflecting specific phonetic or prosodic characteristics without relying on copyrighted works. This opens a potential new research avenue: \textit{To what extent can synthetic songs stand in for real-world ones in improving low-resource \ac{asr}?}

\vspace{-0.3cm}

\section{Conclusion}\label{conclusion}

This study investigated the feasibility of using songs as a novel training resource for Kazakh \ac{asr}. 
The results show that while songs alone do not consistently improve recognition performance and can even degrade generalisation, they provide a meaningful signal when combined with modest corpora such as \ac{cvc} and \ac{fleurs}. 
Mixtures that included songs consistently outperformed single-corpus baselines, with the best results achieved by combining all three datasets (Songs + \ac{cvc} + \ac{fleurs}). 
In this setting, normalised \ac{wer} dropped to 27.6 on \ac{cvc} and 11.8 on \ac{fleurs}, marking substantial improvements over the zero-shot Whisper models and narrowing the gap to the community model trained on the 1,100-hour \ac{ksc} corpus. 
These findings highlight that songs are not a standalone solution but can amplify the value of small existing corpora in low-resource \ac{asr}.

The multi-domain evaluation also revealed important limitations.
Song-based training does not fully transfer to conversational or broadcast speech, and gains remain modest compared to the large-scale upper bound. 
Moreover, orthographic errors (casing, punctuation) remain challenging, though the inclusion of songs helped reduce them in some scenarios, suggesting that lyrics-based data may support better modelling of written-form conventions.

From a practical standpoint, the entire set of fine-tuning experiments cost only \$25 in compute, underscoring that meaningful exploratory research in low-resource \ac{asr} can be conducted with modest resources.

Beyond technical findings, this work raises crucial legal and ethical questions. 
The Songs dataset consists of copyrighted recordings without explicit permission, underscoring the barriers to directly incorporating music into \ac{asr} pipelines. 
This study is therefore intended as a proof of concept rather than a deployable approach. The Kazakh Songs dataset---containing short vocal excerpts (≤ 10 s) aligned with lyric transcriptions---is made available on Hugging Face\footnote{\url{https://huggingface.co/datasets/yeshpanovrustem/kazakh_songs_asr}} under a gated, non-commercial research licence. Full recordings are not distributed; access to excerpts is reviewed and granted solely for academic research purposes.
Future research must explore ethical pathways, including collaboration with artists, the use of public-domain or fair-use excerpts, and the development of frameworks for responsible data use.

Finally, synthetic alternatives represent a promising direction. 
Modern tools such as \texttt{Suno.com} allow the generation of songs with customisable lyrics, genres, and voices, enabling the creation of corpora in low-resource languages without copyright restrictions. 
Such synthetic songs could be designed to cover diverse phonetic contexts, prosodic variations, and stylistic registers, offering a scalable complement to natural data. 
Exploring whether synthetic music can replicate or even surpass the benefits of real songs is a natural next step for advancing low-resource \ac{asr}.

\section{Bibliographical References}\label{sec:refs}
\begingroup
\renewcommand{\section}[2]{}
\bibliographystyle{lrec2026-natbib}
\bibliography{mybib}

@inproceedings{whisper,
    author = {Radford, Alec and Kim, Jong Wook and Xu, Tao and Brockman, Greg and McLeavey, Christine and Sutskever, Ilya},
    title = {{Robust Speech Recognition via Large-Scale Weak Supervision}},
    year = {2023},
    publisher = {JMLR.org},
    booktitle = {Proceedings of the 40th International Conference on Machine Learning},
    articleno = {1182},
    numpages = {27},
    location = {Honolulu, Hawaii, USA},
    series = {ICML'23}
}

@inproceedings{kaznerd,
    title = "{{K}az{NERD}: {K}azakh Named Entity Recognition Dataset}",
    author = "Yeshpanov, Rustem  and
      Khassanov, Yerbolat  and
      Varol, Huseyin Atakan",
    booktitle = "Proceedings of the Thirteenth Language Resources and Evaluation Conference",
    month = jun,
    year = "2022",
    address = "Marseille, France",
    publisher = "European Language Resources Association",
    url = "https://aclanthology.org/2022.lrec-1.44",
    pages = "417--426",
}

@inproceedings{wer,
    author={Andrew Cameron Morris and Viktoria Maier and Phil Green},
    title={{From WER and RIL to MER and WIL}: Improved Evaluation Measures for Connected Speech Recognition},
    year=2004,
    booktitle={Interspeech},
    pages={2765--2768},
    doi={10.21437/Interspeech.2004-668}
}

@inproceedings{cer,
    title={{A Character-level Error Analysis Technique for Evaluating Text Entry Methods}},
    author={MacKenzie, I Scott and Soukoreff, R William},
    booktitle={Proceedings of the second Nordic conference on Human-computer interaction},
    pages={243--246},
    year={2002}
}

@inproceedings{ksc2,
    author={Saida Mussakhojayeva and Yerbolat Khassanov and Huseyin {Atakan Varol}},
    title={{KSC2: An Industrial-Scale Open-Source Kazakh Speech Corpus}},
    year=2022,
    booktitle={Proc. Interspeech 2022},
    pages={1367--1371},
    doi={10.21437/Interspeech.2022-421}
}

@inproceedings{kazakhtts2,
    title = "{{K}azakh{TTS}2: Extending the Open-Source {K}azakh {TTS} Corpus With More Data, Speakers, and Topics}",
    author = "Mussakhojayeva, Saida  and
      Khassanov, Yerbolat  and
      Varol, Huseyin Atakan",
    booktitle = "Proceedings of the Thirteenth Language Resources and Evaluation Conference",
    month = jun,
    year = "2022",
    address = "Marseille, France",
    publisher = "European Language Resources Association",
    url = "https://aclanthology.org/2022.lrec-1.578",
    pages = "5404--5411",
}

@inproceedings{kazemotts,
    title = "{{K}az{E}mo{TTS}: A Dataset for {K}azakh Emotional Text-to-Speech Synthesis}",
    author = "Abilbekov, Adal  and
      Mussakhojayeva, Saida  and
      Yeshpanov, Rustem  and
      Varol, Huseyin Atakan",
    editor = "Calzolari, Nicoletta  and
      Kan, Min-Yen  and
      Hoste, Veronique  and
      Lenci, Alessandro  and
      Sakti, Sakriani  and
      Xue, Nianwen",
    booktitle = "Proceedings of the 2024 Joint International Conference on Computational Linguistics, Language Resources and Evaluation (LREC-COLING 2024)",
    month = may,
    year = "2024",
    address = "Torino, Italia",
    publisher = "ELRA and ICCL",
    url = "https://aclanthology.org/2024.lrec-main.841/",
    pages = "9626--9632"
}

@INPROCEEDINGS{fleurs,
    author={Conneau, Alexis and Ma, Min and Khanuja, Simran and Zhang, Yu and Axelrod, Vera and Dalmia, Siddharth and Riesa, Jason and Rivera, Clara and Bapna, Ankur},
    booktitle={2022 IEEE Spoken Language Technology Workshop (SLT)}, 
    title={{FLEURS: Few-Shot Learning Evaluation of Universal Representations of Speech}}, 
    year={2023},
    volume={},
    number={},
    pages={798-805},
    doi={10.1109/SLT54892.2023.10023141}
}

@inproceedings{cvc,
    title = {{Common Voice: A Massively-Multilingual Speech Corpus}},
    author = {Ardila, Rosana and Branson, Megan and Davis, Kelly and Kohler, Michael and Meyer, Josh and Henretty, Michael and Morais, Reuben and Saunders, Lindsay and Tyers, Francis and Weber, Gregor},
    booktitle = {Language Resources and Evaluation Conference (LREC)},
    month = 05,
    year = {2020},
    address = {Marseille, France},
    publisher = {European Language Resources Association},
    url = {https://aclanthology.org/2020.lrec-1.520},
    pages = {4218--4222},
    language = {English},
    ISBN = {979-10-95546-34-4},
}

@inproceedings{kazakhtts,
    title     = {{KazakhTTS: An Open-Source Kazakh Text-to-Speech Synthesis Dataset}},
    author    = {Saida Mussakhojayeva and Aigerim Janaliyeva and Almas Mirzakhmetov and Yerbolat Khassanov and Huseyin Atakan Varol},
    year      = {2021},
    booktitle = {Interspeech 2021},
    pages     = {2786--2790},
    doi       = {10.21437/Interspeech.2021-2124},
    issn      = {2958-1796},
}

@inproceedings{baas22_interspeech,
    title     = {{Voice Conversion Can Improve ASR in Very Low-Resource Settings}},
    author    = {Matthew Baas and Herman Kamper},
    year      = {2022},
    booktitle = {Interspeech 2022},
    pages     = {3513--3517},
    doi       = {10.21437/Interspeech.2022-112},
    issn      = {2958-1796},
}

@inproceedings{casanova23_interspeech,
    title     = {{ASR data augmentation in low-resource settings using cross-lingual multi-speaker TTS and cross-lingual voice conversion}},
    author    = {Edresson Casanova and Christopher Shulby and Alexander Korolev and Arnaldo Candido Junior and Anderson da Silva Soares and Sandra Aluísio and Moacir Antonelli Ponti},
    year      = {2023},
    booktitle = {Interspeech 2023},
    pages     = {1244--1248},
    doi       = {10.21437/Interspeech.2023-496},
    issn      = {2958-1796},
}

@inproceedings{wang20ia_interspeech,
    title     = {{Improving Cross-Lingual Transfer Learning for End-to-End Speech Recognition with Speech Translation}},
    author    = {Changhan Wang and Juan Pino and Jiatao Gu},
    year      = {2020},
    booktitle = {Interspeech 2020},
    pages     = {4731--4735},
    doi       = {10.21437/Interspeech.2020-2955},
    issn      = {2958-1796},
}

@inproceedings{khare21_interspeech,
    title     = {{Low Resource ASR: The Surprising Effectiveness of High Resource Transliteration}},
    author    = {Shreya Khare and Ashish Mittal and Anuj Diwan and Sunita Sarawagi and Preethi Jyothi and Samarth Bharadwaj},
    year      = {2021},
    booktitle = {Interspeech 2021},
    pages     = {1529--1533},
    doi       = {10.21437/Interspeech.2021-2062},
    issn      = {2958-1796},
}

@article{flores,
    title = {{The {FLoRes}-101 Evaluation Benchmark for Low-Resource and Multilingual Machine Translation}},
    author = "Goyal, Naman  and
      Gao, Cynthia  and
      Chaudhary, Vishrav  and
      Chen, Peng-Jen  and
      Wenzek, Guillaume  and
      Ju, Da  and
      Krishnan, Sanjana  and
      Ranzato, Marc{'}Aurelio  and
      Guzm{\'a}n, Francisco  and
      Fan, Angela",
    editor = "Roark, Brian  and
      Nenkova, Ani",
    journal = "Transactions of the Association for Computational Linguistics",
    volume = "10",
    year = "2022",
    address = "Cambridge, MA",
    publisher = "MIT Press",
    url = "https://aclanthology.org/2022.tacl-1.30/",
    doi = "10.1162/tacl_a_00474",
    pages = "522--538"
}

@inproceedings{veitsman-hartmann-2025-recent,
    title = {{Recent Advancements and Challenges of {T}urkic {C}entral {A}sian Language Processing}},
    author = "Veitsman, Yana  and
      Hartmann, Mareike",
    editor = "Hettiarachchi, Hansi  and
      Ranasinghe, Tharindu  and
      Rayson, Paul  and
      Mitkov, Ruslan  and
      Gaber, Mohamed  and
      Premasiri, Damith  and
      Tan, Fiona Anting  and
      Uyangodage, Lasitha",
    booktitle = "Proceedings of the First Workshop on Language Models for Low-Resource Languages",
    month = jan,
    year = "2025",
    address = "Abu Dhabi, United Arab Emirates",
    publisher = "Association for Computational Linguistics",
    url = "https://aclanthology.org/2025.loreslm-1.25/",
    pages = "309--324"
}

@inproceedings{parcollet25_interspeech,
  title     = {{Loquacious Set: 25,000 Hours of Transcribed and Diverse English Speech Recognition Data for Research and Commercial Use}},
  author    = {Titouan Parcollet and Yuan Tseng and Shucong Zhang and Rogier C. {van Dalen}},
  year      = {2025},
  booktitle = {{Interspeech 2025}},
  pages     = {4053--4057},
  doi       = {10.21437/Interspeech.2025-720},
  issn      = {2958-1796},
}

@article{Zhang2021WENETSPEECHA1,
  title={{WENETSPEECH: A 10000+ Hours Multi-Domain Mandarin Corpus for Speech Recognition}},
  author={Binbin Zhang and Hang Lv and Pengcheng Guo and Qijie Shao and Chao Yang and Lei Xie and Xin Xu and Hui Bu and Xiaoyu Chen and Chenchen Zeng and Di Wu and Zhendong Peng},
  journal={ICASSP 2022 - 2022 IEEE International Conference on Acoustics, Speech and Signal Processing (ICASSP)},
  year={2021},
  pages={6182-6186},
  url={https://api.semanticscholar.org/CorpusID:238419629}
}
\endgroup

\end{document}